\magnification=\magstep1
\tolerance500
\rightline{TAUP-2489-98}
\rightline{April 21, 1998}
\centerline{\bf Description of Unstable Systems}
\centerline{\bf in Relativistic Quantum Mechanics}
\centerline{{\bf in the Lax-Phillips Theory}\footnote{*}{ Presented at
the First International Conference on Parametrized Relativistic
Quantum Theory, PRQT'98, Houston, Feb. 9-11, 1998}}
\bigskip
\centerline{L.P. Horwitz\footnote{\ddag}{Also at
Department of Physics, Bar Ilan University, Ramat Gan, Israel}
 and Y. Strauss }
\centerline{School of Physics, Tel Aviv University}
\centerline{Raymond and Beverly Sackler Faculty of Exact Sciences}
\centerline{Ramat Aviv, 69978 Israel}
\bigskip
\noindent
{\it Abstract:\/} We discuss some of the
experimental motivation for the need for semigroup decay laws,
and  the quantum Lax-Phillips theory of
 scattering and unstable systems.  In this framework, the decay of an
unstable system is described by a semigroup.
 The spectrum of the
generator of the semigroup corresponds to the singularities of the
Lax-Phillips $S$-matrix. In the case of discrete (complex) spectrum of
the generator of the semigroup,
associated with resonances, the decay
law is exactly exponential.  The states corresponding to these
resonances (eigenfunctions of the generator of the semigroup)
lie in the Lax-Phillips Hilbert space, and therefore all physical
properties of the resonant states can be computed.  We show that the
parametrized relativistic quantum theory is a natural setting for the
realization of the Lax-Phillips theory.
 \bigskip
\noindent
{\bf 1. Introduction}
\smallskip
\par There has been considerable effort in recent years in the development
of the theoretical framework of the  scattering theory of Lax and
Phillips$^1$
for the description of quantum mechanical systems$^{2,3,4}$. This work
is motivated by the requirement that the decay law of a decaying system
should be exactly exponential if the simple idea that a set of independent
unstable systems consists of a population for which each element has
a probability, say $\Gamma$, to decay, per unit time. The resulting
exponential law ($\propto e^{-\Gamma t}$) corresponds to an exact semigroup
 evolution of the
state in the underlying Hilbert space, defined as a family of bounded
operators on that space satisfying
$$ Z(t_1) Z(t_2)= Z(t_1+t_2), \eqno(1.1)$$
where $t_1,\, t_2 \geq 0$, and $Z(t)$ may have no inverse. If the
decay of an unstable system is to be associated with an irreversible
process, then its evolution necessarily has the property $(1.1)$.$^5$
The standard model of Wigner and Weisskopf$^6$,
 based on the computation of the survival
amplitude $A(t)$ as the scalar product
$$ A(t) = (\psi, e^{-iHt} \psi) \eqno(1.2)$$
 where $\psi$ is the initial state of the unstable system and $H$ is
the Hamiltonian for the full evolution, results in a good
approximation to an exponential decay law for values of $t$
sufficiently large
(Wigner and Weisskopf$^6$ calculated an atomic linewidth in this
way) but cannot result in a semigroup$^7$.
When applied to a two-channel system, such as the decay of the $K^0$
meson, one easily sees that the poles of the resolvent for the Wigner-Weisskopf
evolution of the two channel systems result in non-orthogonal residues
that generate interference terms, which destroy the semigroup property,
 to accumulate in the calculation of predictions for
regeneration experiments$^8$.
The Yang-Wu$^9$ parametrization of the $K^0$ decay processes, based on a
Gamow$^{10}$ type evolution generated by an effective $2x2$ non-Hermitian
matrix Hamiltonian, on the other hand, results in an evolution that is an
exact semigroup. It appears that the phenomenological parametrization
of refs. 9, which results in semigroup evolution, is indeed
consistent to a high degree of accuracy with the experimental
results on $K$-meson decay$^{11}$.
\par The quantum Lax-Phillips theory provides a framework for
understanding the decay of an unstable system as an irreversible
process.  It appears, in fact,  that this framework is categorical for
the description of irreversible process.
\par The scattering theory of Lax and Phillips assumes the existence
of a Hilbert space $\overline{\cal H}$ of physical states in
which there are two
 distinguished orthogonal subspaces ${\cal D}_+$ and ${\cal D}_-$ with the
 properties
$$\eqalign{ U(\tau)\,{\cal  D}_+ &\subset \,{\cal D}_+  \qquad \tau > 0 \cr
U(\tau) \,{\cal D}_- &\subset \,{\cal D}_- \qquad \tau < 0\cr
\bigcap_\tau \,U(\tau)\,{\cal D}_\pm &= \{0\} \cr
\overline{\bigcup_\tau \,U(\tau)\,{\cal D}_\pm }
&= \overline {\cal H}, \cr} \eqno(1.3)$$
i.e., the subspaces ${\cal D}_\pm$ are stable under the action of the full
 unitary dynamical evolution $U(\tau)$, a function of the physical
 laboratory time, for positive and negatives
times $\tau$ respectively; over all $\tau$, the evolution operator
generates a dense
 set in $\overline{\cal H}$ from either ${\cal D}_+$ or ${\cal D}_-$.
 We shall call ${\cal D}_+$ the {\it outgoing subspace} and
${\cal D}_-$ the {\it incoming subspace} with respect to the group $U(\tau)$.
\par A theorem of Sinai$^{14}$ then assures that $\overline {\cal H}$ can
be represented as a family of Hilbert spaces obtained by foliating
$\overline {\cal H}$ along the real line, which we shall call
$\{s\}$, in the form of a direct integral
$$\overline{\cal H} = \int_\oplus {\cal H}_s,   \eqno(1.4)$$
where the set of auxiliary Hilbert spaces ${\cal H}_s$ are all
isomorphic.  Representing these spaces in terms of square-integrable
functions, we define the norm in the direct integral space (we use
Lesbesgue measure) as
$$ \Vert f \Vert^2 = \int_{-\infty}^\infty \,ds \Vert f_s \Vert^2_H,
\eqno(1.5)$$
where $f \in \overline{H}$ represents
$\overline {\cal H} $ in terms of  the $L^2$ function space
$L^2(-\infty,\infty, H)$,
and $f_s \in H$, the $L^2$ function space
representing ${\cal H}_s$ for any $s$.  We see that, in the framework
of the relativistic quantum theory$^{15}$, the evolution
parameter $\tau$ corresponds to the invariant Stueckelberg evolution
of the system, and the foliation asserted by Sinai$^{14}$ is constructed
within the measure space of the relativistic quantum theory Hilbert space
at each $\tau$. The space $\overline{\cal H}$ then corresponds to the usual
(Stueckelberg)  Hilbert space of states of a relativistic system; it
 therefore provides a natural setting for the realization of the
Lax-Phillips theory.
\par  The Sinai theorem
 furthermore asserts that
there are representations for which the action of the full
evolution group $U(\tau)$ on $L^2(-\infty,\infty, H)$ is translation
by $\tau$ units. Given $D_\pm$ (the $L^2$ spaces representing ${\cal D}_\pm$),
 there is such a representation, called the
{\it incoming representation}$^1$, for which functions in $D_-$ have support
in $L^2(-\infty,0, H)$, and another called the {\it outgoing
representation}, for which functions in $D_+$ have support in $L^2(0,
 \infty,H)$.
\par  Lax and Phillips$^1$ show that there are unitary operators
$W_\pm$,
 called
 wave
operators, which map elements in $\overline{\cal H}$,
respectively, to these representations.  They define an $S$-matrix,
$$ S= W_+W_-^{-1} \eqno(1.6)$$
which connects these representations; it is unitary, commutes with
translations, and maps $L^2(-\infty,0)$ into itself.  The
singularities of this $S$-matrix, in what we shall define as the {\it
spectral representation}, correspond to the spectrum of the generator
of the exact semigroup characterizing the evolution of the unstable
system.
\par With the assumptions stated above on the properties of the
subspaces ${\cal D}_+$ and ${\cal D}_-$, Lax and Phillips$^1$ prove that the
 family
of operators
$$ Z(\tau) \equiv P_+ U(\tau) P_- \qquad (\tau \geq 0), \eqno(1.7)$$
where $P_\pm$ are projections into the orthogonal complements of
${\cal D}_\pm$, respectively, is a contractive, continuous, semigroup.  This
 operator
 annihilates vectors in ${\cal D}_\pm$ and carries the space
$$ {\cal K} = \overline {\cal H} \ominus {\cal D}_+ \ominus
 {\cal D}_- \eqno(1.8)$$
into itself, with norm tending to zero for every element in
${\cal K}$.\footnote{\ddag}{It follows from $(1.7)$ and the stability of
${\cal D}_\pm$
 that $Z(\tau) = P_{\cal K}U(\tau)P_{\cal K}$ as well.}
\par  Functions in the  space ${\overline H}$, representing the
elements of  ${\overline{\cal H}}$,  depend on the variable $s$ as
well as the variables of the auxiliary space $H$. In the
nonrelativistic theory, the  measure space
 of this Hilbert space of states is one dimension larger than that of a
quantum theory represented in the auxiliary space alone.  Identifying
this additional variable with an {\it observable} (in the sense of a
 quantum mechanical observable) time, we may understand this
representation of a state in the nonrelativistic theory
 as a {\it virtual history}. The collection
of such histories forms a quantum ensemble; the absolute square of the
wave function corresponds to the probability that the system
would be found, as a result of measurement, at time $s$
 in a particular configuration in the auxiliary space (in the
state described by this wave function), i.e., an element of
one of the virtual histories. This corresponds precisely to the
interpetation of the relativistic wave functions of the Stueckelberg theory.
The variable $s$, foliating the space according to such virtual
histories, plays a conditional constraint role on the coordinatization of the
spacetime
configurations of the system.
\bigskip
{\bf 2. The Subspaces ${\cal D}_\pm$, Representations,
and the Lax-Phillips $S$-Matrix}
\smallskip
 \par The one-parameter unitary group $U(\tau)$ which acts on the
 Hilbert space
$\overline{\cal H}$ is generated by the invariant operator $K$ which is the
generator of  dynamical evolution of the physical states in $\overline{\cal
H}$; we assume that there exist {\it wave operators} $\Omega_\pm$ which
intertwine this dynamical operator with an unperturbed dynamical
operator
$K_0$.$^{16}$
 We shall assume that $K_0$ has only absolutely
continuous spectrum in $(-\infty,\, \infty)$.
\par We begin the development of the quantum Lax-Phillips theory with
the construction of these representations. In this way, we shall
construct explicitly the foliations described in Section 1.
\par The natural association of
the time variable of the relativistic quantum theory with the
 foliation asserted by the theorem of
Sinai$^{14}$  does not
 correspond to the proper embedding of the relativistic quantum theory
into the Lax-Phillips framework. It is, in fact, clear that for a many
body system, one cannot single out a $t$ variable associated with a
single particle; moreover, the time variable associated with the center
of mass of a system$^{15}$, when it is well-defined, is not conjugate to the
evolution operator ($i$ times its commutator with the evolution operator is the
total $E/M$ of the system), and it is therefore not a candidate
either.  We solve this problem by constructing a foliation on the {\it
free translation representation} of $K_0$.
 The free spectral representation of $K_0$ is defined by
$$   _f\langle \sigma \beta \vert K_0 \vert g \rangle = \sigma\,_f\langle
\sigma \beta \vert g \rangle , \eqno(2.1)$$
This equation corresponds to the Stueckelberg-Schr\"odinger equation$^{15}$ in
$\tau$-independent form, when $\vert g \rangle \rightarrow \vert x
\rangle$; in this case, $K_0$ acting to the right becomes $1/2M$ times
the d'Alembertian (or a sum of such operators for a many body system).
 The solution of the free Stueckelberg problem
therefore provides the transformation function between the {\it model
representation}, for which the spectral values of $x$ have their usual
interpretation as observables in the laboratory, to the free spectral
representation.
Here,  $\vert g \rangle$ is a general element of $\overline{\cal H}$ and
$\beta$ corresponds to the variables (measure space) of the auxiliary
space associated
to each value of $\sigma$, which, with $\sigma$, comprise a complete
spectral set. These constitute the complement in the measure space of
the spectrum of $K_0$.  We shall discuss the structure of this space
in more detail elsewhere.
  The functions $_f\langle \sigma \beta
\vert g \rangle$ may be thought of as a set of functions of the
variables $\beta$ indexed on the variable
 $\sigma$ in a continuous sequence of auxiliary  Hilbert spaces
isomorphic to $H$ .
\par We now proceed to define the incoming and outgoing subspaces
$\cal D_\pm$.  To do this, we define the Fourier transform from
representations according to the spectrum $\sigma$ to the foliation
variable $s$ of $(1.5)$, i.e.,
$$ _f\langle s \beta \vert g \rangle =  \int e^{i\sigma s}\  _f\langle
\sigma \beta \vert g \rangle d\sigma . \eqno(2.2)$$
 Clearly, $K_0$  acts as
the generator of translations in this representation. We shall say
that the set of functions $ _f\langle s \beta \vert g \rangle $ are in
the {\it free translation representation}.
\par  Let us consider
the sets of functions (dense) with support in $L^2(0, \infty)$ and in
$L^2(-\infty,0)$, and call these subspaces $ D_0^\pm$.  The
Fourier transform back to the free spectral representation provides
the two sets of Hardy class functions
$$ _f\langle \sigma \beta \vert g_0^\pm \rangle =  \int e^{-i\sigma
s}\
 _f\langle s \beta \vert g_0^\pm \rangle ds  \in H_\pm , \eqno(2.3)$$
for $g_0^\pm \in D_0^\pm$.
\par We may now define the subspaces ${\cal D}_\pm$ in the Hilbert space of
states ${\overline{\cal H}}$.  To do this we first map these Hardy
class functions in ${\overline H}$ to ${\overline{\cal H}}$, i.e., we
define the subspaces ${\cal D}_0^{\pm}$ by
$$\int \sum_\beta \vert\sigma \beta \rangle_f \  _f\langle \sigma
\beta \vert g_0^\pm \rangle d\sigma \in {\cal D}_0^\pm.  \eqno(2.4)$$
\par As remarked above, we  assume that there are wave operators
which intertwine $K_0$
with the full evolution $K$, i.e., that the limits
$$ \lim_{\tau \to \pm \infty} e^{iK\tau} e^{-iK_0 \tau} = \,
\Omega_\pm \eqno(2.5)$$
exist on a dense set in ${\overline{\cal H}}$.
\par  The construction of ${\cal D}_\pm $ is then completed with the help
of the wave operators.  We define these subspaces by
$$\eqalign{ {\cal D}_+ &= \Omega_+ {\cal D}_0^+  \cr
{\cal D}_- &= \Omega_- {\cal D}_0^-  .\cr} \eqno(2.6)$$
We remark that these subspaces are not produced by the same unitary
map. This procedure is necessary to realize the Lax-Phillips structure
non-trivially. If a single unitary map were used, then there would
exist a transformation into the space of functions on $L^2(-\infty,
\infty, H)$ which has the property that all functions with support on
the positive half-line represent elements of ${\cal D}_+$, and all
functions with support on the negative half-line represent elements of
${\cal D}_-$ in the same representation; the resulting Lax-Phillips
$S$-matrix would then be trivial.
 The requirement that ${\cal D}_+$
and ${\cal D}_-$ be orthogonal is not an immediate consequence of our
construction; as we shall see, this result is associated with the
analyticity of the operator which corresponds to the Lax-Phillips
 $S$-matrix.
\par In the following, we construct the Lax-Phillips $S$-matrix and
the Lax-Phillips wave operators.
\par The wave operators defined by $(2.5)$ intertwine $K$ and $K_0$, i.e.,
$$ K \Omega_\pm  = \Omega_\pm K_0.        \eqno(2.7)$$
We may therefore construct the outgoing (incoming) spectral representations
from the free spectral representation.  Since
$$ \eqalign{K\Omega_\pm \vert \sigma \beta \rangle_f &= \Omega_\pm K_0 \vert
\sigma \beta \rangle_f \cr
&=\sigma \Omega _\pm \vert \sigma \beta \rangle_f,\cr} \eqno(2.8)$$
we may identify
$$ \vert \sigma \beta \rangle_{out \atop in} = \Omega_\pm \vert \sigma
\beta \rangle_f . \eqno(2.9)$$
Let us now act on these functions with the Lax-Phillips $S$-matrix in the
free spectral representation, and require the result to be the {\it outgoing}
representer of $g$:
$$ \eqalign{{_{out}\langle} \sigma \beta \vert g)
&= {_f\langle} \sigma \beta \vert \Omega_+^{-1} g) \cr
&=\, \int d\sigma'\, \sum_{\beta'}{_f\langle} \sigma \beta \vert
 {\bf S}\vert \sigma'\beta' \rangle_f \,\,
{_f\langle}\sigma' \beta' \vert \Omega_-^{-1} g) \cr} \eqno(2.10)$$
where ${\bf S}$ is the Lax-Phillips $S$-operator
 (defined on ${\overline{\cal H}}$).
Transforming the kernel to the free translation representation
with the help of
$(2.2)$, i.e.,
$$ {_f\langle} s \beta\vert {\bf S} \vert s' \beta' \rangle_f =
{1 \over (2\pi)^2}
\int d\sigma d\sigma' \, e^{i\sigma s} e^{-i\sigma's'}
{_f\langle} \sigma \beta \vert
 {\bf S}\vert \sigma'\beta' \rangle_f , \eqno(2.11)$$
we see that the relation  $(2.10)$ becomes, after using the Fourier
transform in a similar way to
transform the {\it in} and {\it out } spectral representations to
the corresponding {\it in} and {\it out} translation representations,
$$\eqalign{ {_{out}\langle} s\beta \vert g) = {_f\langle} s\beta
 \vert \Omega_+^{-1} g) &=
\int ds'\, \sum_{\beta'} {_f\langle} s \beta\vert {\bf S} \vert s'
\beta' \rangle_f
\,{_f\langle} s' \beta' \vert \Omega_-^{-1} g) \cr
&=  \int ds'\, \sum_{\beta'} {_f\langle} s \beta\vert {\bf S} \vert s' \beta'
 \rangle_f {_{in}\langle} s'\beta' \vert g). \cr}  \eqno(2.12)$$
Hence the Lax-Phillips $S$-matrix is given by
$$ S= \{ {_f\langle} s \beta\vert {\bf S} \vert s' \beta' \rangle_f
\},
 \eqno(2.13)$$
 in free translation representation. It follows from the intertwining
property $(2.7)$ that
$$ {_f\langle} \sigma \beta \vert{\bf S}\vert \sigma' \beta' \rangle_f =
\delta(\sigma - \sigma') S^{\beta \beta'}(\sigma). \eqno(2.14)$$
 \bigskip
\noindent
{\bf 3. Orthogonality of ${\cal D}_\pm$ and Analyticity of the
$S$-Matrix.}
\par The orthogonality of ${\cal D}_\pm $ follows from the analytic
properties of the $S$-matrix.  To display this analyticity property,
 we study the operator ${\bf S}$ in the form (from $(2.10)$; this operator
coincides with the relativistic quantum mechanical $S$-matrix)
$$ {\bf S} = \Omega_+^{-1}\Omega_- = \lim_{\tau \rightarrow \infty}
e^{iK_0\tau} e^{-2iK\tau} e^{iK_0\tau}. \eqno(3.1)$$
It then follows in the standard way$^{17}$ that
$${_f\langle} \sigma\beta \vert {\bf S}
 \vert \sigma' \beta' \rangle_f =
 \delta(\sigma- \sigma')\{ \delta^{\beta\beta'} - 2\pi i
 \,{_f\langle}\sigma\beta
\vert{\bf T} (\sigma
+ i\epsilon) \vert \sigma \beta' \rangle_f \},  \eqno(3.2)$$
where
$$ {\bf T}(z) = V + VG(z)V = V + VG_0 {\bf T}(z). \eqno(3.3)$$
We remark that, by this construction, we see that
$S^{\beta\beta'}(\sigma)$
 is {\it analytic in the upper half plane} in $\sigma$.
\par We have constructed the incoming and outgoing subspaces
 ${\cal D}_\pm$ in $(2.6)$.  It is essential for application
 of the Lax-Phillips theory that these subspaces be orthogonal,
 i.e., for every $f_+ \in {\cal D}_+, \, f_- \in {\cal D}_-,$ that
 $(f_+, f_-) = 0$.  If
$$ \eqalign{ f_+ &= \Omega_+ f_0^+ \cr
f_- &= \Omega_- f_0^- , \cr} \eqno(3.4)$$
mapped from functions in ${\cal D}_0^\pm$, we see that the orthogonality
condition is
$$ (f_+, f_-) = (f_0^+, \Omega_+^{-1} \Omega_- f_0^-) = 0. \eqno(3.5)$$
  As shown in $(2.11)$,
 the $S$-matrix in free representation transforms the incoming to
 the outgoing representation; we may therefore write the scalar
product in
 $(3.5)$ as
$$(f_+, f_-) = \sum_{\beta \beta'} \int ds ds' \, (f_0^+ \vert s
 \beta \rangle_{out} \,\,{_f\langle} s \beta \vert {\bf S} \vert s'
 \beta' \rangle_f \,\, {_{in}\langle} s' \beta' \vert f_0^- )  \eqno(3.6)$$
Now, in the free translation representation, we have
$$\eqalign{{_f\langle} s \beta \vert {\bf S} \vert s' \beta' \rangle_f &=
     \int d\sigma d\sigma' \, e^{i\sigma s} e^{-i\sigma' s'}
     {_f\langle}\sigma \beta \vert {\bf S} \vert \sigma' \beta' \rangle_f \cr
&= \int d\sigma e^{i\sigma(s-s')}  S^{\beta \beta'}(\sigma) \cr
&= S^{\beta \beta'} (s-s').  \cr } \eqno(3.7)$$
The  function $S(\sigma)^{\beta \beta'}$ is analytic in the upper
 half plane; it may have a null co-space, but is otherwise regular.
  Its singularity lies in the lower half plane .
  To find a non-vanishing value for  $S^{\beta \beta'}(s-s')$,
 we must close the contour in the lower half plane.
 This can only be done if $s' >s$.  For $s' <s$,
 one must close in the upper half plane, and there
 $S(\sigma)$ has no singularity, so the integral vanishes.  Hence
$S^{\beta \beta'}(s-s')$ takes ${\cal D}_-$ to ${\cal D}_-$
 in the incoming representation, and the subspaces ${\cal D}_+$
 and ${\cal D}_-$ are orthogonal.

\bigskip
\noindent
{\bf 4. Conclusions and Discussion}
\smallskip
\par  It was shown that a necessary condition for a non-trivial
 Lax-Phillips theory, for which the singularities of
 the $S$-matrix in the spectral variable constitute the spectrum
 of the generator of the semigroup, is that the evolution operator
 act as a smooth (operator-valued) integral kernel on the time axis
 in the free translation representation.$^4$ We have shown in this paper
 that a  {\it pointwise} (in  spacetime $x$) dynamical evolution operator in
 what we have called the model representation, in which the
 Hamiltonian of a system and its spacetime variables appear with
 their usual laboratory interpretation, maps into a smooth,
 non-trivial kernel (through $(2.1)$) in the free translation
 representation, and
 therefore satisfies this necessary condition.  The
relativistic quantum theory therefore provides a natural framework for the
Lax-Phillips theory.
\bigskip
\noindent
\smallskip
{\it Acknowledgements}
\par One of us (L.H.) wishes to thank S.L. Adler, E. Eisenberg, C. Newton,
 C. Piron and T.T. Wu for discussions on the Lax-Phillips theory,
 and I. Dunietz , Y.B. Hsiung, and B. Winstein
 and other colleagues at the Fermilab for many discussions of the
 implications of the current experiments on $K$ meson decay as well
 as potential applications in $B$ physics
for the theoretical structure that we have described here.
 \bigskip
\noindent
\frenchspacing
{\bf References}
\item{1.} P.D. Lax and R.S. Phillips, {\it Scattering Theory},
Academic Press,
 N.Y. (1967).
\item{2.} C. Flesia and C. Piron, Helv. Phys. Acta {\bf 57}, 697
(1984).
\item{3.} L.P. Horwitz and C. Piron, Helv. Phys. Acta {\bf 66}, 694 (1993).
\item{4.} E. Eisenberg and L.P. Horwitz, in {\it Advances in
 Chemical Physics}, {\bf XCIX}, p. 245, ed. I. Prigogine and
 S. Rice, John Wiley and Sons, N.Y. (1997).
\item{5.} C. Piron, {\it Foundations of Quantum Physics},
 Benjamin/Cummings, Reading, Mass. (1976).
\item{6.} V.F. Weisskopf and E.P. Wigner, Zeits. f. Phys.
 {\bf 63}, 54 (1930); {\bf 65}, 18 (1930).
\item{7.} L.P. Horwitz, J.P. Marchand and J. LaVita,
 J. Math. Phys. {\bf 12}, 2537 (1971); D. Williams,
 Comm. Math. Phys. {\bf 21}, 314 (1971).
\item{8.} L.P. Horwitz and L. Mizrachi, Nuovo Cimento {\bf 21A}, 625 (1974).
\item{9.} T.D. Lee, R. Oehme and C.N. Yang, Phys. Rev.
 {\bf 106}, 340 (1957); T.T. Wu and C.N. Yang, Phys. Rev. Lett. {\bf
13}, 380
 (1964).
\item{10.}G. Gamow, Z. Phys. {\bf 51}, 204(1928).
\item{11.} B. Winstein, {\it et al}, {\it Results from the
 Neutral Kaon Program at Fermilab's Meson Center Beamline, 1985-1997},
 FERMILAB-Pub-97/087-E, published on behalf of the E731, E773
 and E799 Collaborations, Fermi National Accelerator Laboratory,
 P.O. Box 500, Batavia, Illinois 60510.
\item{12.} S.R Wilkinson, C.F. Bharucha, M.C. Fischer, K.W. Madison,
P.R. Morrow, Q. Niu, B. Sundaram, and M. Raizen, Nature {\bf 387}, 575
(1997)
\item{13.}  W. Baumgartel, Math. Nachr. {\bf 69}, 107 (1975);
 L.P. Horwitz and I.M. Sigal, Helv. Phys. Acta
 {\bf 51}, 685 (1978); G. Parravicini, V. Gorini
 and E.C.G. Sudarshan, J. Math. Phys. {\bf 21}, 2208
 (1980); A. Bohm, {\it Quantum Mechanics: Foundations
 and Applications\/,} Springer, Berlin (1986); A. Bohm,  M. Gadella
and
 G.B. Mainland, Am. J. Phys. {\bf 57}, 1105 (1989); T. Bailey and
 W.C. Schieve, Nuovo Cimento {\bf 47A}, 231 (1978).
\item{14.} I.P. Cornfield, S.V. Formin and Ya. G. Sinai,
{\it Ergodic Theory}, Springer, Berlin (1982).
\item{15.} E.C.G. Stueckelberg, Helv. Phys. Acta {\bf 14}, 372,
588(1941); {\bf 15}, 23 (1942); L.P. Horwitz and C. Piron,
Helv. Phys. Acta {\bf 48}, 316 (1974); R.E. Collins and J.R. Fanchi,
Nuovo Cimento {\bf 48A}, 314 (1978); J.R. Fanchi, {\it Paramterized
Relativistic Quantum Theory}, Kluwer, Norwell, Mass. (1993), and
references therein.
\item{16.} L.P. Horwitz and A. Soffer, Helv. Phys. Acta {\bf 53}, 112 (1980).
\item{17.} For example, J.R. Taylor, {\it Scattering Theory},
 John Wiley and Sons,
 N.Y. (1972); R.J. Newton, {\it Scattering Theory of Particles
 and Waves}, McGraw Hill, N.Y. (1976).

\vfill
\end
\bye